# Magneto optical rotation in a GaAs Quantum Well Waveguide

**Ali Mortezapour[1*], Mohsen Ghaderi Goran Abad[2] and Mohammad Mahmoudi[2]**

[1] Department of Physics, University of Guilan, P. O. Box 41335–19141, Rasht, Iran

[2] Physics Department, University of Zanjan, P.O. Box 45195-313, Zanjan, Iran

Corresponding author E-mail: mortezapour@guilan.ac.ir

**Abstract:**

The interaction of two orthogonally polarized beams and a four-level GaAs quantum well (QW) waveguide is investigated. It is shown that, by applying a static magnetic field normal to the propagation direction of the driving beams, the birefringence can be induced in the QW waveguide. Moreover, it is demonstrated that the dephasing rate between two ground states of the QW waveguide makes it a dichromatic medium and can also diminish the induced birefringence. Our results show how a large and complete magneto-optical rotation in the QW waveguide can be obtained via adjusting the intensity of the magnetic field and also the length of the QW waveguide.

## 1- Introduction

Magneto-optical rotation (MOR) is a fundamental and practical phenomenon originating from the interaction between light and matter. In the MOR, the polarization plane of the linearly polarized light rotates during its passage through the medium placed in a static magnetic field. The mechanism behind the MOR is related to the different complex refractive indexes value for two circular component of linearly polarized light which is induced by a static magnetic field. Referring to the light propagation direction, the magnetic field can be applied in two different directions. They can be either parallel or perpendicular to the propagation direction of light. If the magnetic field is applied parallel to the light propagation direction, such phenomenon is called the Faraday effect[1], whereas for a perpendicular applied magnetic field, the phenomenon is known as the Voigt effect (in gases)[2] or the Cotton-Mouton effect (in liquids) [3]. The MOR in the atomic gases was initially reported by Macaluso and Corbino [4, 5]. They observed that the MOR depends strongly on light frequency. In the last few years, control and enhancement of the MOR in atomic gases have been well studied both theoretically and experimentally [6-11]. However, it was shown that the MOR of the linearly polarized beam can be enhanced via quantum interference between two spontaneous emissions in V-type three-level atomic system [12]. In a distinguished study, the polarization rotation of the linearly polarized Gaussian pulse is analyzed during its propagation through the cold atomic gas in the presence of a longitudinal magnetic field [13]. The Phase-controlled optical Faraday rotation in a closed-loop atomic system was recently studied and it was shown that the rotation of polarization plane can be controlled by the intensity and relative phase of applied fields [14].

Today, various applications are known for the MOR. One of the earliest applications of the MOR goes back to the polarization spectroscopy [15]. The polarization spectroscopy is a spectroscopic technique based on polarization properties of light. Where, the changes in the transmitted light polarization (relative to the incidental one) allows us to infer the media properties. Furthermore, the MOR brings out the most sensitive methods of measuring magnetic fields. This so-called high-sensitivity optical magnetometry [16, 17] uses the polarization of the light to measure the Zeeman shift of the atomic sublevels and then the magnetic field's strength. It has been recognized that the MOR plays an important role in measuring parity violation (PV) in heavy atomics vapors caused by the weak interaction [18]. Parity nonconservation occurs when in a transition the products of the parity quantum numbers of the initial and final sub-systems are not identical. It is shown that in Thallium Vapor, the PV signature would be a frequency shift in the Zeeman sublevels splitting. Hence, the PV affects the rotation angle of the polarization plane of the polarized light in a MOR experiment [19].

In spite of the fact that the atomic systems are ideal for controlling and manipulating of optical properties, semiconductor nanostructure devices such as quantum dots and quantum wells (QWs) are well promising for practical applications and profiting straightly from the recent advances in micro- and nano-technologies. Many optical properties of semiconductor QWs are similar to the atomic system. However, it is well known that QWs have some superiority to the atoms due to their intrinsic advantages such as large electro-dipole moments and high nonlinear optical coefficients. In addition, QWs transition energies and symmetries can be constructed from the selected materials with a high degree of accuracy. Therefore such advantages make it possible to see many physical effects at room temperature. Among lots of engineered QWs, Gallium Arsenside (GaAs) QW waveguide is one of the most important ones that has wide applications in optoelectronic devices. Up to now, several typical phenomena such as electromagnetically induced transparency (EIT) [20], four wave mixing [21], optical bistability [22, 23], efficient weak-light amplification[24], slow optical soliton pairs [25], generation of frequency entangled states [26] electromagnetically induced grating [27], have been studied in a GaAs QW waveguide.

Motivated by these, in this paper, we consider a GaAs QW waveguide, interacting coherently with a linearly polarized ($\sigma$-polarized) probe beam and a π-polarized control beam. As well as the QW waveguide is positioned in a transverse magnetic field. It is demonstrated that by appropriate adjusting the intensity of the magnetic field and also the length of the QW waveguide, a complete and large MOR can be observed for the linearly polarized probe beam during its passage through the QW waveguide.

This paper arrangement is as following: in section 2, we present our model and obtain density matrix equations describing the evolution of the system and explain mathematical equations governing the MOR. In section 3, we discuss, in details, the numerical and the analytical results of the absorption, dispersion and the MOR of the probe beam in the QW waveguide. Finally, the main conclusions of the work are outlined in Section 4.

**2-Model and equations**

Consider a four-level semiconductor GaAs quantum well (QW) waveguide (WG) that is placed in a transverse magnetic field as shown in Fig. 1(a). In such a semiconductor structure, there are two degenerate conduction band states with $S_z = 1/2$ and $S_z = -1/2$ ($|3\rangle, |4\rangle$) and two degenerate light-hole (LH) valance band states with $J_z = \pm 1/2$ ($|1\rangle, |2\rangle$). As depicted in Fig. 1(b), applying a static magnetic field $\vec{B} = -B\hat{y}$ transverse to the plane of the QW (Voigt geometry), leads to Zeeman splitting in both the conduction and the LH bands but does not affect the dipole selection rule [28]. The splitting in the conduction and the LH bands are denoted by $2\Delta_B = -2g_s\mu_B B/\hbar$ and $2\Delta_{lh} = -2g_J\mu_B B/\hbar$ respectively. Where $\mu_B$ is the Bohr magneton, $B$ is the magnetic field strength in the $y$ direction, $g_s$ and $g_J$ are the effective Landé factors of the conduction band and the LH band respectively. It is noteworthy that, with the appropriate choice of the well barrier material in making up the GaAs QWs, a negative value can be obtained for $g_s$. While $g_J$, for the GaAs, is always a negative number [29]. In this work, like as references [20-26], we consider only the interband dipole optical transitions between the doubly degenerate LH valence bands and the doubly degenerate conduction bands. Although, for a more realistic description it is important to take into account the influence of the interactions between carriers and also that of scatterings of the carriers by phonons and disorders. However, the present model can give a qualitative illustration which is conceptually useful and explains some of the key features.

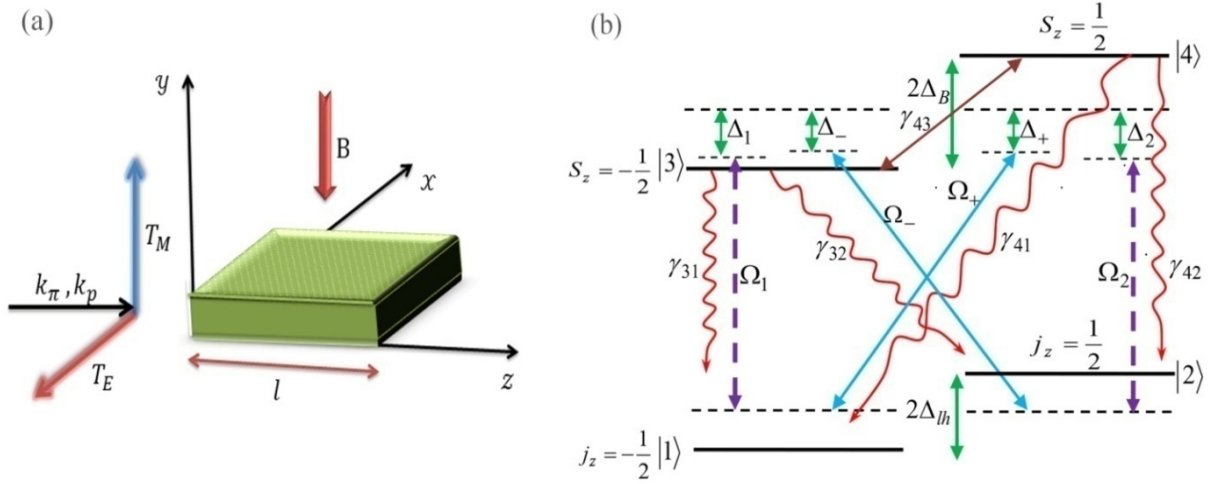

Fig. 1

In this way, as shown in Fig. 1(a), a linearly polarized (TE-polarized) probe beam $\vec{E}_p(z,t) = \hat{e}_x E_p \exp[-i(\omega_p t - k_p z)] + c.c$ with frequency $\omega_p$ and wave vector $k_p$ normal to the magnetic field direction is applied to the QW waveguide. It is well known that the linearly polarized field can be regarded as a linear combination of the left- and the right-circularly polarized components. According to the polarization selection rule, the left-circularly polarized component ($\sigma_-$) of the probe beam with Rabi frequency $\Omega_- = (\vec{\mu}_{32} \cdot \hat{e}_-) E_- / \hbar$ excites the $|3\rangle \leftrightarrow |2\rangle$ transition with resonant frequency $\omega_{32}$, while the right-circularly polarized component ($\sigma_+$) with Rabi frequency $\Omega_+ = (\vec{\mu}_{41} \cdot \hat{e}_+) E_+ / \hbar$ excites the $|4\rangle \leftrightarrow |1\rangle$ transition with resonant frequency $\omega_{41}$. Note that $E_+ = E_- = E_p / \sqrt{2}$ and $\vec{\mu}_{32} = \vec{\mu}_{41}$, so we have $\Omega_+ = \Omega_-$ [24, 30]. At the same time, a $\pi$-polarized (TM-polarized) strong control beam $\vec{E}_c(z,t) = \hat{e}_y E_\pi \exp[-i(\omega_c t - k_c z + \phi)] + c.c$ drives the transitions $|3\rangle \leftrightarrow |1\rangle$ (with resonant frequency $\omega_{31}$) and $|4\rangle \leftrightarrow |2\rangle$ (with resonant frequency $\omega_{42}$) with Rabi frequencies $\Omega_1 = (\vec{\mu}_{31} \cdot \hat{e}_y) E_\pi / \hbar$ and $\Omega_2 = (\vec{\mu}_{42} \cdot \hat{e}_y) E_\pi / \hbar$, respectively. For the reason that $\vec{\mu}_{42} = -\vec{\mu}_{31}$, $\Omega_1$ and $\Omega_2$ are in the opposite signs ($\Omega_2 = -\Omega_1$) [20, 26]. Note that $\vec{\mu}_{mn}$ denotes the dipole moment for the transition $|m\rangle \leftrightarrow |n\rangle$ and $\hat{e}_i (i = +, -, y)$ characterizes the polarization unit vector of the electric field and parameter $\phi$ represents the relative phase of the driving beams.

In the interaction picture, the Hamiltonian describing the dynamics of the system in the dipole and rotating-wave approximations is given by:

$$V_I = -\hbar\Omega_1 e^{-i[(\Delta_1+\Delta_B-\Delta_{lh})t-k_c z+\phi]}|3\rangle\langle 1| - \hbar\Omega_2 e^{-i[(\Delta_2-\Delta_B+\Delta_{lh})t-k_c z+\phi]}|4\rangle\langle 2|$$
$$-\hbar\Omega_+ e^{-i[(\Delta_+-\Delta_B-\Delta_{lh})t-k_p z]}|4\rangle\langle 1| - \hbar\Omega_- e^{-i[(\Delta_-+\Delta_B+\Delta_{lh})t-k_p z]}|3\rangle\langle 2| + c.c. \quad (1)$$

Here $\Delta_+ = \omega_p - \omega_{41}$ and $\Delta_- = \omega_p - \omega_{32}$ represent the detuning of the two circular components of the probe beams and $\Delta_1 = \omega_c - \omega_{31}$ and $\Delta_2 = \omega_c - \omega_{42}$ are the detuning of the control beam relative to the transitions $|3\rangle \leftrightarrow |1\rangle$ and $|4\rangle \leftrightarrow |2\rangle$ respectively. For the sake of simplicity, we would assume that $\hbar = 1$. Note that the LH band states and conduction band states are degenerate ($\omega_{41} = \omega_{32}$), so we have; $\Delta_+ = \Delta_- = \Delta_p$ and $\Delta_1 = \Delta_2 = \Delta_\pi$. From Eq. (1), it is straightforward to attain the following Liouville equations for the density matrix elements:

$$\dot{\rho}_{11} = \gamma_{31}\rho_{33} + \gamma_{41}\rho_{44} + i\Omega_1 e^{i\phi}\rho_{31} - i\Omega_1 e^{-i\phi}\rho_{13} + i\Omega_+\rho_{41} - i\Omega_+\rho_{14},$$
$$\dot{\rho}_{22} = \gamma_{32}\rho_{33} + \gamma_{42}\rho_{44} + i\Omega_-\rho_{32} - i\Omega_-\rho_{23} + i\Omega_2 e^{i\phi}\rho_{42} - i\Omega_2 e^{-i\phi}\rho_{24},$$
$$\dot{\rho}_{33} = -(\gamma_{31}+\gamma_{32})\rho_{33} + i\Omega_1 e^{-i\phi}\rho_{13} - i\Omega_1 e^{i\phi}\rho_{31} + i\Omega_-\rho_{23} - i\Omega_-\rho_{32},$$
$$\dot{\rho}_{21} = -[\Gamma_{21}/2 + i(\Delta_1 - \Delta_- - 2\Delta_{lh})]\rho_{21} + i\Omega_-\rho_{31} + i\Omega_2 e^{i\phi}\rho_{41} - i\Omega_+\rho_{24} - i\Omega_1 e^{-i\phi}\rho_{23},$$
$$\dot{\rho}_{31} = -[\Gamma_{31}/2 - i(\Delta_1 + \Delta_B - \Delta_{lh})]\rho_{31} + i\Omega_1 e^{-i\phi}(\rho_{11} - \rho_{33}) + i\Omega_-\rho_{21} - i\Omega_+\rho_{34},$$
$$\dot{\rho}_{41} = -[\Gamma_{41}/2 - i(\Delta_+ - \Delta_B - \Delta_{lh})]\rho_{41} + i\Omega_+(\rho_{11} - \rho_{44}) + i\Omega_2 e^{-i\phi}\rho_{21} - i\Omega_1 e^{-i\phi}\rho_{43}, \quad (2)$$
$$\dot{\rho}_{32} = -[\Gamma_{32}/2 - i(\Delta_- + \Delta_B + \Delta_{lh})]\rho_{32} + i\Omega_-(\rho_{22} - \rho_{33}) + i\Omega_1 e^{-i\phi}\rho_{12} - i\Omega_2 e^{-i\phi}\rho_{34},$$
$$\dot{\rho}_{42} = -[\Gamma_{42}/2 - i(\Delta_2 - \Delta_B + \Delta_{lh})]\rho_{42} + i\Omega_2 e^{-i\phi}(\rho_{22} - \rho_{44}) - i\Omega_-\rho_{43} + i\Omega_+\rho_{12},$$
$$\dot{\rho}_{43} = -[\Gamma_{43}/2 - i(\Delta_2 - \Delta_- - 2\Delta_B)]\rho_{43} - i\Omega_-\rho_{42} - i\Omega_1 e^{i\phi}\rho_{41} + i\Omega_2 e^{-i\phi}\rho_{23} + i\Omega_+\rho_{13},$$
$$\rho_{11} + \rho_{22} + \rho_{33} + \rho_{44} = 1.$$

In the above density matrix equation, the phenomenological added overall dephasing rates $\Gamma_{ij}$ are given by $\Gamma_{21} = \gamma_{21}^d$, $\Gamma_{31} = \gamma_3 + \gamma_{31}^d$ ($\gamma_3 = \gamma_{31} + \gamma_{32}$), $\Gamma_{32} = \gamma_3 + \gamma_{32}^d$, $\Gamma_{41} = \gamma_4 + \gamma_{41}^d$ ($\gamma_4 = \gamma_{41} + \gamma_{42}$), $\Gamma_{42} = \gamma_4 + \gamma_{42}^d$, $\Gamma_{43} = \gamma_3 + \gamma_4 + \gamma_{43}^d$. Where, as shown in Fig. 1(b), the decay rates $\gamma_{31}$ and $\gamma_{41}$ ($\gamma_{32}$ and $\gamma_{42}$) are the decay rates of two conduction bands $|3\rangle$ and $|4\rangle$ to the state $|1\rangle$ ($|2\rangle$) respectively. Furthermore $\gamma_{ij}^d$ is the dephasing decay rate of the quantum coherence of the $|i\rangle \leftrightarrow |j\rangle$ transition. Note that the dephasing decay rate between two conduction band states ($\gamma_{43}^d$) is the decay rate for the spin coherence. It is worth mentioning that in Eq. (2), we assumed that the multi-photon resonant condition $\Delta_+ + \Delta_- = \Delta_1 + \Delta_2$ ($\Delta_p = \Delta_\pi$) to be implemented throughout the calculations. The susceptibility corresponding to right(left) circular polarization of the probe beam is defined as $\chi^+$ ($\chi^-$) which can be written in terms of dimensionless quantities as follow:

$$\chi_{\pm} = \left(\frac{\alpha}{4\pi k_p}\right) S_{\pm} \qquad (3)$$

Where $S_{\pm}$, are the normalized susceptibilities, are given by

$$S_+ = \frac{\rho_{41}\gamma_{41}}{\Omega_+}, \qquad S_- = \frac{\rho_{32}\gamma_{32}}{\Omega_-}, \qquad (4)$$

Note that the imaginary and real parts of $S_{\pm}$ represent the absorptions and the dispersions of the two circular components of the probe field respectively. Here the quantity $\alpha l = 4\pi k_p l N |\mu|^2 / \hbar\gamma$ is the laser field absorption at the line center. Parameter $N$ denotes effective density averaged over the cross section of the probe beam and $l$ is the length of the waveguide and $|\vec{\mu}_{32}| = |\vec{\mu}_{41}| = |\mu|$. Here we have also assumed $\gamma_{32} = \gamma_{41} = \gamma$ for simplicity.

It is considered that probe beam is polarized along the $\hat{x}$ axis and propagates in the $\hat{z}$ direction. To measure the rotation of the polarization direction of the probe beam, the intensity of the y-polarization component of the output probe beam is calculated. Experimentally it is measured by crossing the output probe beam through a y-polarized analyzer. The intensity of the y-polarization component of the output probe beam with respect to the intensity of the incident probe beam is given by [13]

$$T_y = \frac{|(E_{p(out)})_y|^2}{|E_{p(int)}|^2} = \frac{1}{4} |\exp[i\alpha l S_+/2] - \exp[i\alpha l S_-/2]|^2. \qquad (5)$$

The normalizing output intensity in the $\hat{x}$ direction with respect to the input intensity is also given by

$$T_x = \frac{|(E_{p(out)})_x|^2}{|E_{p(int)}|^2} = \frac{1}{4} |\exp[i\alpha l S_+/2] + \exp[i\alpha l S_-/2]|^2. \qquad (6)$$

It is clear that when the real(imaginary) parts of the normalized susceptibilities of two circularly components of the beam are different $\text{Re}[S_+] \neq \text{Re}[S_-]$ ( $\text{Im}[S_+] \neq \text{Im}[S_-]$ ), the QW waveguide is a birefringent (dichromatic) medium. Both of these two mechanisms can rotate the polarization plane of the probe beam. However, when $\text{Re}[S_+] \neq \text{Re}[S_-]$ and $\text{Im}[S_+] \approx \text{Im}[S_-] = \beta$, we can write $T_x$ and $T_y$ as follow:

$$T_x = \frac{e^{-\alpha l \beta}}{4} \left| \exp[i\alpha l \, \text{Re}[S_+]/2] + \exp[i\alpha l \, \text{Re}[S_-]/2] \right|^2,$$

$$T_y = \frac{e^{-\alpha l \beta}}{4} \left| \exp[i\alpha l \, \text{Re}[S_+]/2] - \exp[i\alpha l \, \text{Re}[S_-]/2] \right|^2.$$
(7)

In this circumstance, there are generally two desirable situations that are interesting to us. The first one occurs when the value of $\beta$ is negative. Thus the probe beam attains the gain and then will be amplified in the output. The second situation occurs when the value of $\beta$ is positive, where $\alpha l \beta \ll 1$. Under such circumstances, the attenuation of the intensity of the probe beam does not remarkably occur during passage through the medium. Note that in the both above situations, the rotation is merely due to the birefringence.

## 3- Result and discussion

In the following, we solve the density matrix equations (2) numerically in the steady state condition and use equation (4) to get the results for the MOR under the assumption of specific parameters. The results of our calculations and the conclusions are then discussed. We indicate that the results of our present study may be realized at low temperature (up to 4 K) [22-24]. Throughout this paper it is assumed that $\gamma_{32} = \gamma_{41} = \gamma$, $\gamma_{31} = \gamma_{42} = \gamma'$, $\Delta_{lh} = 0$, $\Delta_p = \Delta_\pi = \Delta$, $\gamma_{31}^d = \gamma_{32}^d = \gamma_{41}^d = \gamma_{42}^d = 0$. For a typical QW waveguide, we have $\gamma = 10^{11} Hz$. In the following all the other parameters are scaled by $\gamma$, chosen as the unit. Fig.2 illustrates the evolution of the imaginary ((a), (b)) and real ((c), (d)) parts of $S_+$ and $S_-$ and their difference versus detuning of the applied beams ($\Delta$) in the absence (column (I)) and in the presence (column (II)) of the dephasing rate between LH states ($\Gamma_{21}$). The values of the parameters are taken as follow: $\phi = 0$, $\Delta_B = 9\gamma$, $\Omega_+ = \Omega_- = 1\gamma$, $\Omega_1 = -\Omega_2 = 1\gamma$, $\gamma_{43}^d = 0.05\gamma$, $\gamma' = 0.01\gamma$ for $\Gamma_{21} = 0$ (column (I)) and $\Gamma_{21} = 0.05\gamma$ (column (II)). In the absence of $\Gamma_{21}$, It is seen that the absorption of the right and left circular components of the linearly polarized probe beam are exactly the same. Then their difference is zero. Whereas, as depicted in Fig. 2(c), the dispersion of these components are completely different from each other. Moreover, it is seen that the maximum difference occurs when the driving fields are on resonance ($\Delta = 0$). This implies that in the absence of $\Gamma_{21}$, the QW waveguide shows the birefringence behavior. Thus the MOR rises solely due to the birefringence induced in the QW waveguide.

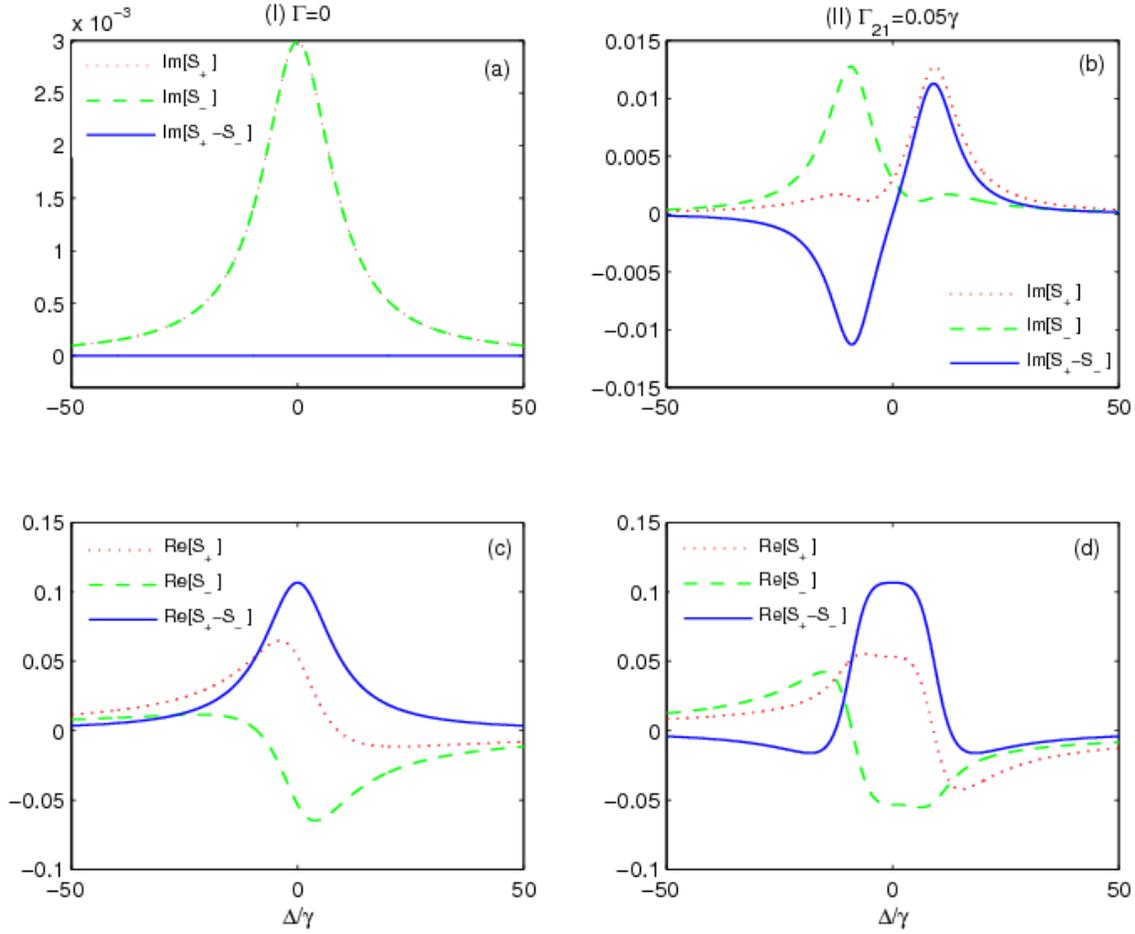

Fig. 2

However, in the presence of $\Gamma_{21}$ one can see that the absorption of two circular components of the probe beam are different except $\Delta = 0$. As well as it can be observed in Fig. 2(d), these components experience different dispersions except for two values of $\Delta$. Moreover, a close look at this figure reveals that the maximum difference between the dispersions of circular components of the probe beam occurs around $\Delta = 0$. Thus we find that in the presence of $\Gamma_{21}$ the QW waveguide just at $\Delta = 0$ behaves as a pure birefringent medium. While for the other values of $\Delta$, the dichroism appears in the QW waveguide.

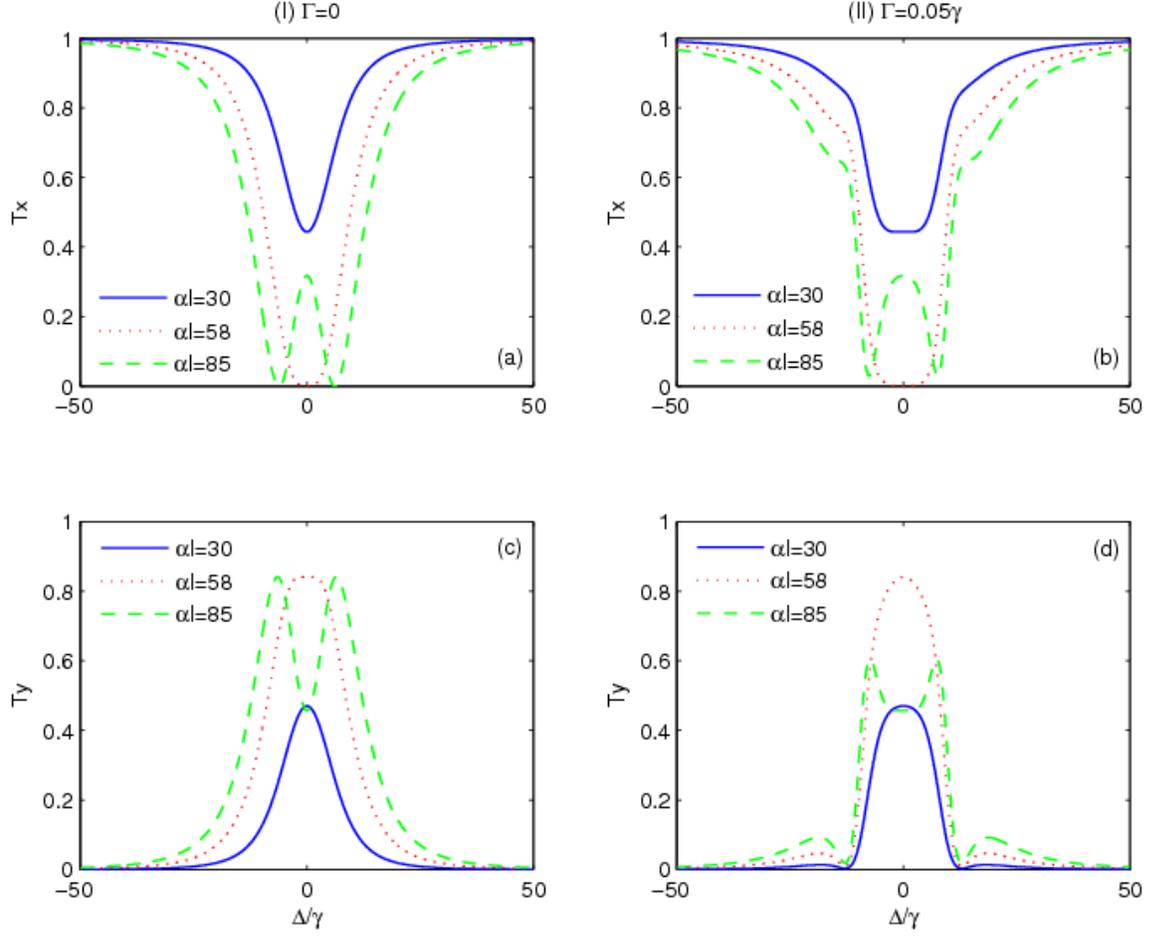

Fig. 3

We are now in a position to discuss about the MOR in the QW waveguide. To this goal, we plot the transmitted intensity in the $\hat{x}$ and $\hat{y}$ directions ($T_x$ and $T_y$) versus detuning of the driving beams for $\alpha l =30$ (Solid blue line), $\alpha l = 58$ (Dotted red line) and $\alpha l = 85$ (Dashed green line) in Fig. 3. The other parameters are the same as those used in Fig. 2. Fig. 3 illustrates obviously that the MOR is much dependent on $\alpha l$. It is seen that for $\alpha l$ =30, either in the absence or in the presence of $\Gamma_{21}$, there is not any complete polarization rotation of the probe beam. However, for the case of $\alpha l$ =58, a complete MOR of the probe beam occurs at $\Delta = 0$, in which the QW waveguide behaves as a pure birefringent medium. While the width of the window for the complete MOR is decreased in the presence of $\Gamma_{21}$. Regarding $\alpha l$ =85, the complete MOR occurs just in the absence of $\Gamma_{21}$ for the two values of $\Delta$, where the polarization rotation is solely due to the birefringence induced in the QW waveguide. Meanwhile, in the presence of $\Gamma_{21}$, a great deal of fraction of the intensity of the probe beam is wasted inside the QW waveguide.

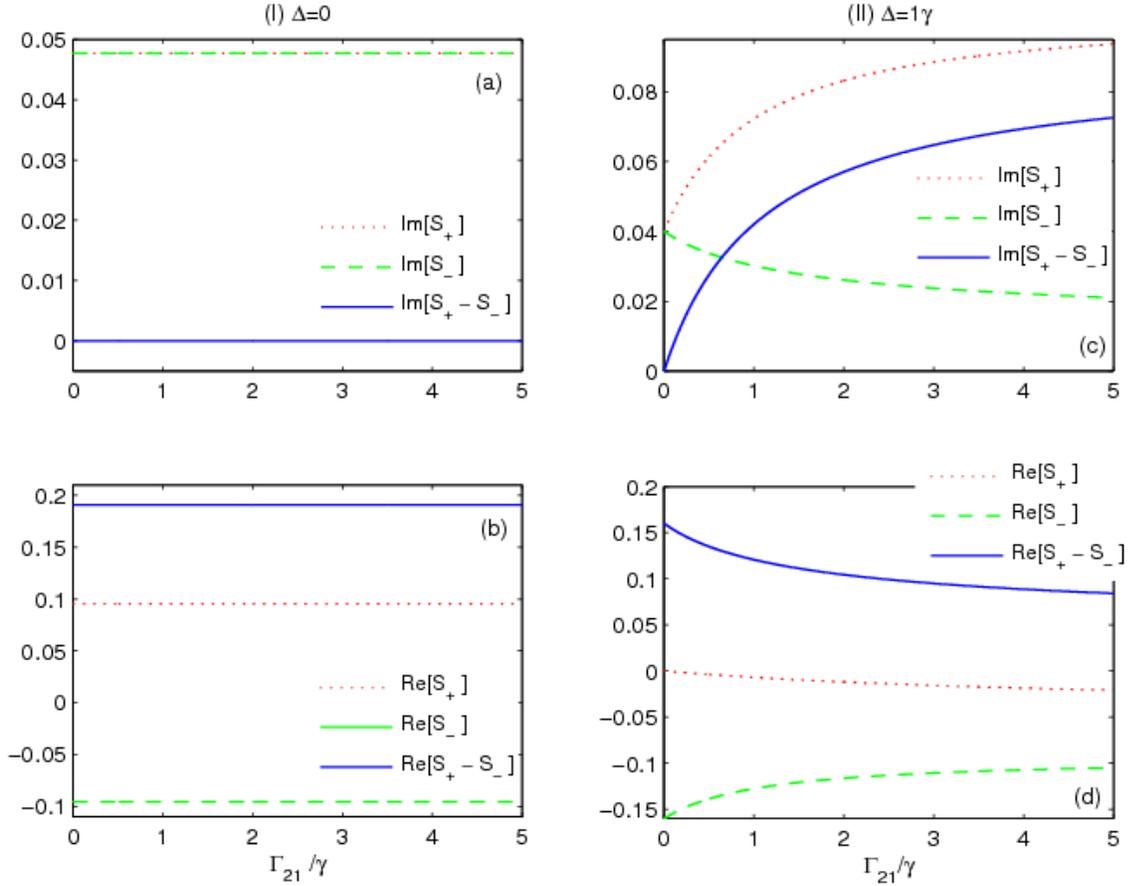

Fig. 4

To explore further the role of $\Gamma_{21}$, the imaginary ((a), (b)) and real ((c), (d)) parts of $S_+$ and $S_-$ and their difference as a functions of $\Gamma_{21}$ are plotted in Fig. 4. The parameters are taken as $\Delta_B = 1\gamma$, $\Delta = 0$ for column (I) and $\Delta = 1\gamma$ for column (II). The others are the same as those used in Fig. 2. An investigation of Fig. 4 shows that for the resonance situation of the driving beams, $\Gamma_{21}$ does not have any effect on the absorption and the dispersion of two circular components of the probe beam. However, in the case of $\Delta = 1\gamma$, emersion of $\Gamma_{21}$ induces the dichroism in the QW waveguide. Moreover, increasing of this parameter results in strengthening of the dichroism and weakening of the birefringence simultaneously.

Now, we study the effect of the relative phase of the driving beams on the MOR. Fig. 5 displays the imaginary (a) and real (b) parts of $S_+$ (dotted), $S_-$ (dashed) and their difference (solid) and also $T_x$ and $T_y$ versus the relative phase of the driving beams ($\phi$). The values of parameters are taken as $\Delta = 0$, $\Delta_B = 5\gamma$, $\alpha l = 30$ and the others are the same as those used in Fig. 2.

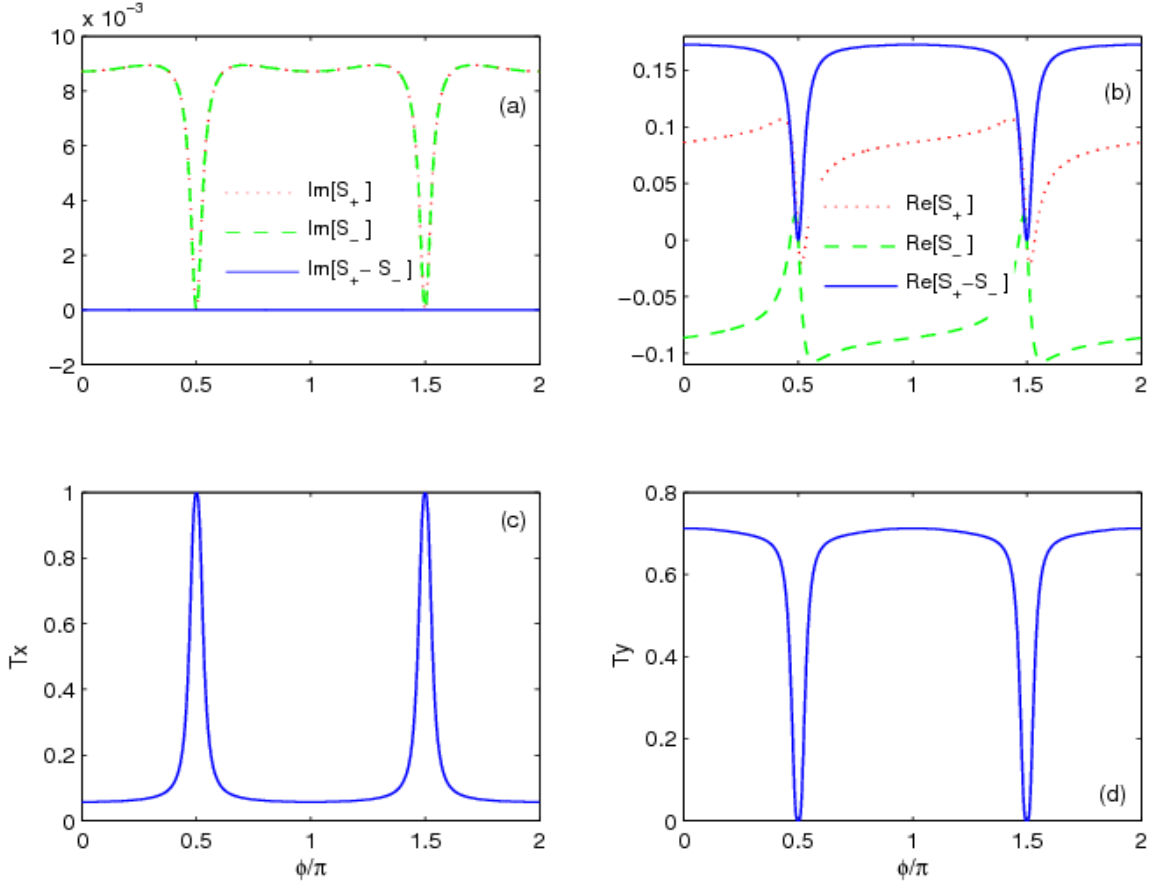

Fig. 5

It is seen that in the case of $\phi = \pi/2$ and $\phi = 3\pi/2$, the absorption and the dispersion of two circular components of the probe beam are equal to zero (see Fig.5 (a) and (b)). It means that the QW waveguide behaves as a transparent medium and the induced birefringence disappears in the whole process. Thus, the MOR cannot occur for $\phi = \pi/2$ and $\phi = 3\pi/2$ (see Fig.5 (c) and (d)). Our calculations show that the results of Fig. 5 are the same for all values of $\Delta_B$ and $\alpha l$.

Under the steady state regime and when the parameters are taken as $\gamma_{31} = \gamma_{42} = \gamma'$, $\phi = 0$, $\Delta = 0$, $\gamma_{31}^d = \gamma_{32}^d = \gamma_{41}^d = \gamma_{42}^d = \gamma_{43}^d = \gamma_{21}^d = 0$, $\Omega_+ = \Omega_- = \Omega$ and $\Omega_1 = -\Omega_2 = \Omega_\pi$. The exact analytic solution of the density matrix equations (Eq.(2)) results in following equations for $\rho_{32}$ and $\rho_{41}$:

$$\rho_{32} = \frac{-2\Delta_B + i(\gamma + \gamma')}{(\gamma + \gamma')^2 + 8(\Omega^2 + \Omega_\pi^2) + 4\Delta_B^2}\Omega, \qquad \rho_{41} = \frac{2\Delta_B + i(\gamma + \gamma')}{(\gamma + \gamma')^2 + 8(\Omega^2 + \Omega_\pi^2) + 4\Delta_B^2}\Omega. \qquad (8)$$

Then, using equations (4), the normalized susceptibilities ($S_\pm$) are given by

$$S_- = \frac{-2\Delta_B + i(\gamma + \gamma')}{(\gamma + \gamma')^2 + 8(\Omega^2 + \Omega_\pi^2) + 4\Delta_B^2}\gamma, \qquad S_+ = \frac{2\Delta_B + i(\gamma + \gamma')}{(\gamma + \gamma')^2 + 8(\Omega^2 + \Omega_\pi^2) + 4\Delta_B^2}\gamma. \qquad (9)$$

As depicted in equation (9) the imaginary parts of $S_+$ and $S_-$ are the same. Thus by assuming $\text{Im}[S_-] = \text{Im}[S_+] = \beta$, it is straightforward to get the output intensity in the $\hat{x}$ and $\hat{y}$ directions and the rotation angle as follow:

$$T_x = \frac{\exp[-\alpha l \beta]}{2}(1 + Cos[\frac{2\alpha l \Delta_B}{(\gamma + \gamma')^2 + 8(\Omega^2 + \Omega_\pi^2) + 4\Delta_B^2}\gamma]), \qquad (10)$$

$$T_y = \frac{\exp[-\alpha l \beta]}{2}(1 - Cos[\frac{2\alpha l \Delta_B}{(\gamma + \gamma')^2 + 8(\Omega^2 + \Omega_\pi^2) + 4\Delta_B^2}\gamma]), \qquad (11)$$

$$\Phi = \tan^{-1}[\sqrt{T_y/T_x}], \qquad (12)$$

where $\beta = \dfrac{i(\gamma + \gamma')}{[\gamma + \gamma']^2 + 8(\Omega^2 + \Omega_\pi^2) + 4\Delta_B^2}\gamma$.

In the rest of the paper, we keep our discussion by using the analytical results of Eqs. (9)-(11).

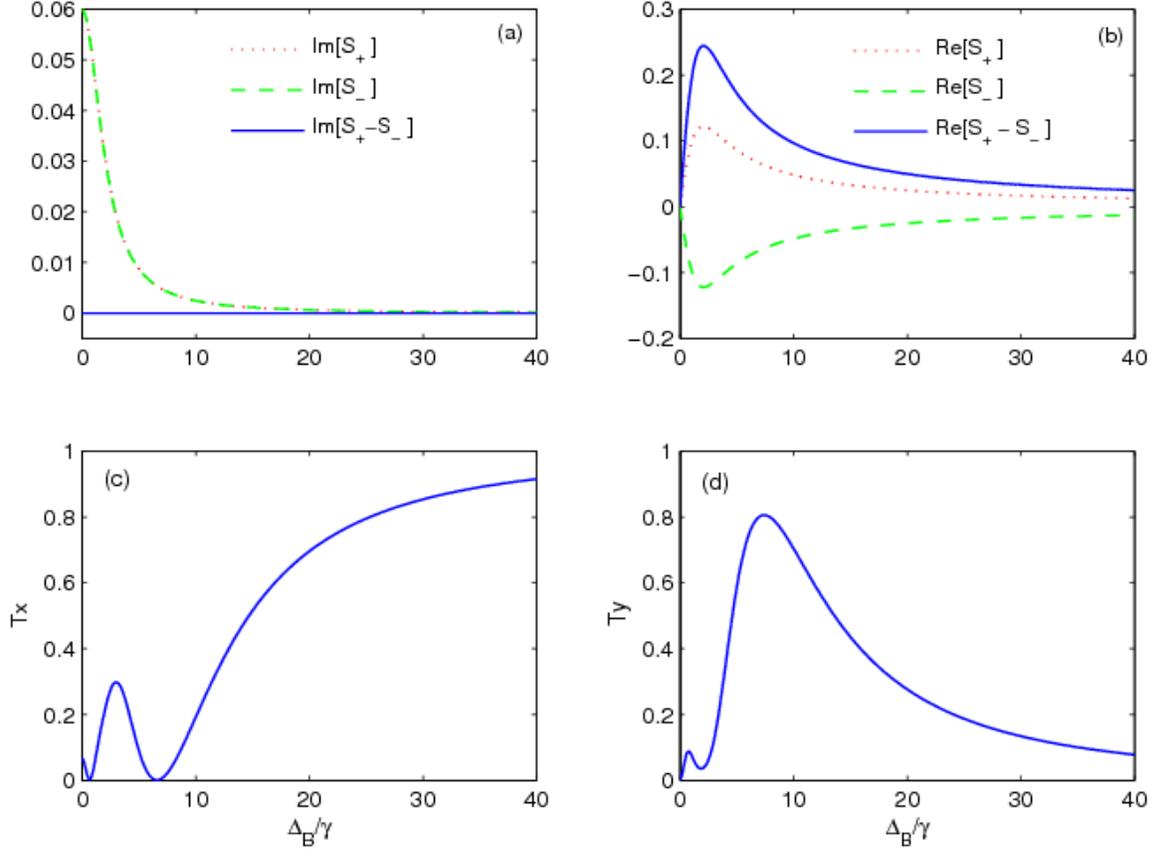

Fig. 6

In Fig. 6 we plot the imaginary (a) and real (b) parts of $S_+$ (dotted), $S_-$ (dashed) and their difference (solid) and also $T_x$ and $T_y$ versus $\Delta_B$. The values of the parameters are chosen as follow: $\phi = 0$, $\Omega_\pi = \Omega = 1\gamma$, $\gamma' = 0.01\gamma$, $\Gamma_{21} = 0$, $\alpha l = 45$. It is seen that the increasing of $\Delta_B$ leads to the reducing of the absorption of the two circular components of the probe beam. Nevertheless, the dispersion of them immediately begins to be different after the magnetic field is switched on. That is to say, the applied magnetic field induces the birefringence in the QW waveguide. Further, from Fig. 6(b), one can see that the maximum birefringence occurs at $\Delta_B = 2\gamma$. Also for the case that $\Delta_B = 2\gamma$ the birefringence grows fast for $\Delta_B \leq 2\gamma$ and reduces smoothly for $\Delta_B > 2\gamma$. On the other hand, Fig. 6(c) and 6(d) show that a complete and large MOR of the probe beam occurs at $\Delta_B = 7\gamma$. That is due to the fact that all of the output intensity (80 percent of the input intensity) of the probe beam exits in the $\hat{y}$ direction.

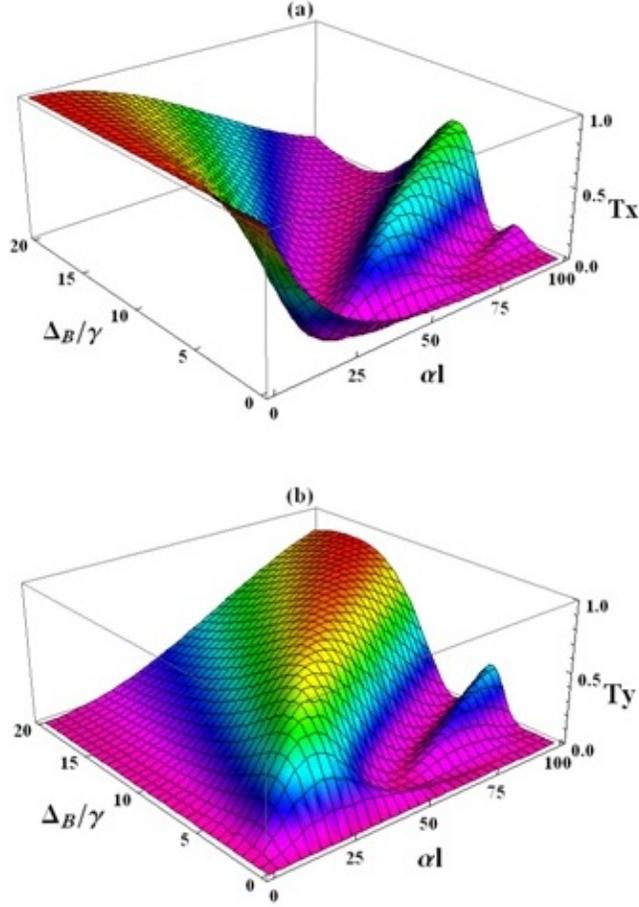

Fig. 7

To better bringing out the roles of the magnetic field and also the length of the QW waveguide on the enhancement of the MOR of the probe beam, we displayed $T_x$ (a) and $T_y$ (b) as functions of $\Delta_B$ and $\alpha l$ in Fig. 7. The other parameters are the same as those used in Fig. 6. Fig. 7 shows that a large MOR can be obtained by a proper choice of the intensity of the magnetic field and also the length of the QW waveguide. It is clear that in the absence of any QW waveguide, the MOR would not happen. Whereas, for every amount of $\alpha l$, there is a special value of $\Delta_B$ in which an enhancement of $T_y$ is achievable.

Finally, we briefly address the requirement of the magnetic field and $\alpha l$. The ranges of the Zemman splitting of $\Delta_B < 10\gamma$, which can generate large and complete MOR, can be caused by the magnetic field of the strength $B < 10T$. It is noteworthy that such a high magnetic field strength can be created by a superconducting magnet at sufficiently low temperature. In the case of $\alpha l$, the range of $\alpha l \leq 100$ correspond to the length of $l \leq 1 \mu m$ for the GaAs QW waveguide, which is reliable. Although, it is possible to control the value of $\alpha l$ by changing the electronic number density ($N$).

**4- Conclusion:**

In conclusion, we have proposed and analyzed a scheme for generating the MOR in a GaAs QW waveguide. This can be achieved by applying a linear-polarized (TE-polarized) probe beam and a $\pi$-polarized (TM-polarized) control beam to the QW waveguide in the presence of a transverse static magnetic field. We have shown that, when the driving beams are on resonance, the magnetic field makes the QW waveguide a pure birefringent medium. However, for the off-resonance situation and just when the dephasing rate between two LH states is present the dichroism appears and the induced birefringence weakens. Moreover, it was deduced that with the appropriate adjusting the intensity of the magnetic field and also the length of the QW waveguide, a large and complete MOR is achievable.

**Acknowledgment**

Ali Mortezapour would like to thank Professor Nai-Hang Kwong for his insightful discussions on this work.

**Figure caption**

**Fig. 1.** (a) Schematic of the magnetic field relative to the QW waveguide. A $\pi$-polarized (TM-polarized) control beam and a $\sigma$-polarized (TE-polarized) probe beam. Probe and control beams propagate in the direction parallel to the QW plane ($\hat{z}$ direction). (b) Energy levels of the GaAs QW waveguide under a transverse magnetic field. Conduction band states are labeled with $S_{\pm} = 1/2$ and the valance band states are labeled with $J_{\pm} = 1/2$.

**Fig. 2.** Imaginary (a) and real (b) parts of $S_+$ (dashed) and $S_-$ (dotted) and their difference (solid) versus detuning of the applied beams. Values of parameters are taken as follow: $\phi = 0$, $\Delta_B = 9\gamma$, $\Omega_\pi = \Omega = \gamma$, $\gamma' = 0.01\gamma$. Column (I) for $\Gamma_{21} = 0$ and column (II) for $\Gamma_{21} = 0.05\gamma$.

**Fig. 3.** Transmitted intensity in the $\hat{x}$ and $\hat{y}$ direction ($T_x$ and $T_y$) versus detuning of the driving beams for $\alpha l = 30$ (Solid blue line), $\alpha l = 58$ (Dotted red line), $\alpha l = 85$ (Dashed green line). Other parameters are the same as those used in Fig. 2.

**Fig. 4.** (a) and (b) absorption and (c) and (d) dispersion of circular components of the probe beam is plotted versus $\Gamma_{21}$. Column (I) for $\Delta = 0$ and column (II) for $\Delta = 1\gamma$. Other parameters are the same as in Fig. 2.

**Fig. 5.** (a) Imaginary and (b) real parts of $S_+$ (dotted), $S_-$ (dashed), their difference and $T_x$ and $T_y$ versus $\phi$. Parameters values of are $\Delta = 0$, $\Delta_B = 5\gamma$, $\Omega_\pi = \Omega = \gamma$, $\gamma' = 0.01\gamma$, $\alpha l = 30$.

**Fig. 6.** (a) Imaginary and (b) real parts of $S_+$ (dotted), $S_-$ (dashed), their difference and $T_x$ and $T_y$ versus $\Delta_B$. Parameters are chosen as: $\phi = 0$, $\Omega_\pi = \Omega = 1\gamma$, $\gamma' = 0.01\gamma$, $\Gamma_{21} = 0$, $\alpha l = 45$.

**Fig. 7.** Plots of (a) $T_x$ and (b) $T_y$ as a functions of $\Delta_B$ and $\alpha l$. Parameters used are same as in Fig. 6.